\documentclass[times,twocolumn,final]{elsarticle}

\usepackage{medima}
\usepackage{framed,multirow}
\usepackage{multirow}%
\usepackage{booktabs}%
\usepackage{subfig}
\usepackage{lipsum}
\usepackage{amsmath} 

\usepackage[export]{adjustbox}
\usepackage{float}
\usepackage{dblfloatfix}

\usepackage{amssymb}
\usepackage{latexsym}

\usepackage{url}

\usepackage{hyperref}

\definecolor{newcolor}{rgb}{.8,.349,.1}

\begin{document}

\verso{M. Imran \textit{et~al.}}

\begin{frontmatter}

\title{CIS-UNet: Multi-Class Segmentation of the Aorta in  Computed Tomography Angiography via Context-Aware Shifted Window Self-Attention}

\author[1]{Muhammad Imran}
\author[2]{Jonathan R Krebs}
\author[3]{Veera Rajasekhar Reddy Gopu}
\author[2]{Brian Fazzone}
\author[31]{Vishal Balaji Sivaraman}
\author[3]{Amarjeet Kumar}
\author[2]{Chelsea Viscardi}
\author[4]{Robert Evans Heithaus}
\author[1]{Benjamin Shickel}
\author[5]{Yuyin Zhou\fnref{fn1}}
\author[2]{Michol A Cooper\fnref{fn1}}
\author[1]{Wei Shao\corref{cor1}\fnref{fn1}}
\cortext[cor1]{Corresponding author. 
  E-mail address: weishao@ufl.edu (W. Shao)}
\fntext[fn1]{Equal contribution as senior authors.}

\address[1]{Department of Medicine, University of Florida, Gainesville, FL 32611, USA}
\address[2]{Department of Surgery, University of Florida, Gainesville, FL 32611, USA}
\address[3]{Department of Computer and Information Science and Engineering, University of Florida, Gainesville, FL 32611, USA}
\address[31]{Department of Electrical and Computer Engineering, University of Florida, Gainesville, FL 32611, USA}
\address[4]{Department of Radiology, University of Florida, Gainesville, FL 32611, USA}
\address[5]{Department of Computer Science and Engineering, University of California, Santa Cruz, CA 95064, USA}


\begin{abstract}
Advancements in medical imaging and endovascular grafting have facilitated minimally invasive treatments for aortic diseases. Accurate 3D segmentation of the aorta and its branches is crucial for interventions, as inaccurate segmentation can lead to erroneous surgical planning and endograft construction. Previous methods simplified aortic segmentation as a binary image segmentation problem, overlooking the necessity of distinguishing between individual aortic branches. In this paper, we introduce Context Infused Swin-UNet (CIS-UNet), a deep learning model designed for multi-class segmentation of the aorta and thirteen aortic branches. Combining the strengths of Convolutional Neural Networks (CNNs) and Swin transformers, CIS-UNet adopts a hierarchical encoder-decoder structure comprising a CNN encoder, symmetric decoder, skip connections, and a novel Context-aware Shifted Window Self-Attention (CSW-SA) as the bottleneck block.
Notably, CSW-SA introduces a unique utilization of the patch merging layer, distinct from conventional Swin transformers. It efficiently condenses the feature map, providing a global spatial context and enhancing performance when applied at the bottleneck layer, offering superior computational efficiency and segmentation accuracy compared to the Swin transformers. We trained our model on computed tomography (CT) scans from 44 patients and tested it on 15 patients. CIS-UNet outperformed the state-of-the-art SwinUNetR segmentation model, which is solely based on Swin transformers, by achieving a superior mean Dice coefficient of 0.713 compared to 0.697, and a mean surface distance of 2.78 mm compared to 3.39 mm. CIS-UNet's superior 3D aortic segmentation offers improved precision and optimization for planning endovascular treatments. Our dataset and code will be publicly available.

\end{abstract}


\end{frontmatter}



\section{Introduction}
\label{sect:introduction}
The aorta is the largest artery of the body, carrying oxygenated blood from the heart to the head, neck, upper extremities, abdomen, pelvis, and lower extremities. 
Pathologies of the aorta and its main branches, like dissection, aneurysm, and atherosclerotic disease, can be immediate threats to life or limb, requiring prompt surgical evaluation and treatment \citep{writing20222022}. 
Advances in medical imaging and therapies, including the introduction of minimally invasive, or ``endovascular'' aortic stent grafts, have led to a paradigm shift in the management of aortic disease  \citep{parodi1991transfemoral}.  
Endovascular abdominal aortic aneurysm repair, for example, is now performed as first-line therapy in over 80\% of patients \citep{chaikof2018society}. 
For minimally invasive repairs involving branch vessels, a detailed 3D analysis of the aortic and branch vessel anatomy is essential. This includes measuring the centerline and diameter of the aorta and individual aortic branches for appropriate device selection.
Current aorta segmentation methods treat the problem as a binary segmentation task, which is inadequate for complex repairs that involve multiple branch vessels. 

High-resolution computed tomography angiography (CTA) is the gold standard imaging modality to assess vascular pathology and facilitate preoperative planning \citep{chaikof2018society,writing20222022}. 
CTA is widely available, rapidly obtained, and can image the entire aorta and branches with high spatial resolution \citep{writing20222022}. 
After obtaining a preoperative CTA, most surgeons rely on commercially available software platforms for 3D planning.  
However, contemporary commercially available 3D software is very costly and semi-automatic and can require significant time and training to obtain the clinically relevant measurements of interest \citep{oderich2009modified}.

While many accurate and fully automated aortic segmentation models have been developed, they primarily focus on binary segmentation of the main aorta and aortic branches. Therefore, these models do not provide individual measurements of the centerline and diameter for different aortic branches, which are critical for complex endovascular treatment planning. Additionally, these methods are limited by their ability to accurately model anatomic variation, vessel tortuosity, and stenotic vessel anatomy. Consequently, segmentation models might demonstrate adequate results with imaging from healthy control aortas but fail under more anatomically complex conditions.
To the best of our knowledge, there is no existing method that specifically targets the multi-class segmentation of the aorta along with its thirteen branches.

In this paper, we present the Context-Infused Swin-UNet (CIS-UNet), a hybrid deep learning framework for multi-class segmentation of the aorta and its branches. CIS-UNet leverages the strengths of both Convolutional Neural Networks (CNNs) and Swin transformers, adopting a hierarchical U-shaped encoder-decoder structure. Despite the efficiency of Swin Transformers' self-attention mechanisms, the local window-based attention imposes limitations on modeling global dependencies. To overcome this, we propose Context-aware Shifted Window Self-Attention (CSW-SA), which condenses the feature map along the spatial dimension and integrates it with the original output from the Swin Transformer block. This process extracts and incorporates global contextual information across self-attention windows. Additionally, we found that using CSW-SA solely in the bottleneck layer enables accurate and detailed segmentation with minimal computation overhead, effectively capturing the complex anatomical structures and variations present in aortic and branch vessel imaging. CIS-UNet allows for enhanced precision in identifying and measuring each vessel, which is crucial for planning and executing advanced endovascular procedures.

This paper has the following major contributions:
\begin{itemize}
    \item 	We curated a large dataset of CT images along with accurate segmentation of the aorta and thirteen aortic branches, which will serve as a valuable resource for future research.
    \item  We proposed a novel context-aware shifted window self-attention block that efficiently segment very subtle, complex, and heterogeneous objects.
    \item Our model outperformed current state-of-the-art in terms of segmentation accuracy and computational efficiency.   
\end{itemize}

\section{Related Work}
\label{sect:related-work}
\subsection{U-Net Based Segmentation Models}
Convolutional neural networks (CNNs) \citep{fukushima1988neocognitron, lecun1998gradient, krizhevsky2012imagenet, simonyan2014very, he2016deep} have been a dominant approach for computer vision and medical image analysis due to their efficiency in extracting local image features. 
U-Net is among the most widely used architectures for medical image segmentation, characterized by its contracting pathway that captures context and an expanding pathway focused on localization \citep{ronneberger2015u}. Despite its effectiveness, deeper U-Net architectures can face the vanishing gradient problem \citep{raza2023dresu}. The Residual U-Net \citep{kerfoot2019left} integrates the classic U-Net architecture with residual connections, enhancing feature learning and addressing the vanishing gradient issue in medical image segmentation. The dResU-Net \citep{raza2023dresu} takes this a step further, combining a deep residual network encoder with a U-Net decoder. This blend incorporates both high- and low-level features, using shortcuts and skip connections for improved image segmentation and faster training. 
Building on this trend of integrating attention mechanisms, Swin UNETR \citep{tang2022self} combines the Swin Transformer \citep{liu2021swin} and U-Net. This model incorporates a hierarchical encoder for self-supervised pre-training, optimized using specific proxy tasks to recognize human anatomy patterns, setting new standards in various medical image segmentation tasks.

\subsection{Prior Work on Aorta Segmentation}
Research on aorta segmentation has typically approached the problem as binary image segmentation. 
These studies do not differentiate between the primary aorta and its various branches. 
Some works emphasize the segmentation of the entire aorta \citep{fantazzini20203d, chen2021multi}, while others focus on specific sections of the aorta \citep{deng2018graph, sedghi2019automated, gu2021fusing, chen2022deep, saitta2022deep}. 
These segmentations are performed on both contrast-enhanced CT scans \citep{fantazzini20203d, bonechi2021segmentation, saitta2022deep, chen2021multi} and non-contrast enhanced CT scans \citep{deng2018graph, sedghi2019automated, gu2021fusing, chen2022deep}. 
The methodologies behind these works range from traditional techniques \citep{deng2018graph, sedghi2019automated} to deep learning-based methods. 
Within the realm of deep learning, some researchers utilize 2D segmentation models either trained individually \citep{bonechi2021segmentation, chen2022deep} or aggregated from multiple views \citep{fantazzini20203d}. 
Others have adopted direct 3D segmentation models \citep{saitta2022deep, chen2021multi} or a fusion of 2D and 3D models \citep{gu2021fusing}.

\subsection{Vision Transformer Based Image Segmentation Methods}
\label{sect:transformer-based-techniques}
Medical image segmentation is challenging due to the necessity of capturing both global and local features within the input image. CNNs are highly effective in preserving local spatial information and generating high-resolution outputs for detailed segmentation, whereas transformers excel in learning long-range dependencies and comprehending global contexts. Vision transformers (ViTs) \citep{dosovitskiy2020image, liu2021swin} have emerged as a significant advancement in computer vision, employing self-attention mechanisms to overcome some limitations of CNNs, such as capturing long-range dependencies. However, they encounter challenges like increased computational costs and difficulties in local feature extraction \citep{guo2022cmt, chen2021visformer}. Addressing these challenges, the Swin Transformer \citep{liu2021swin} utilizes shifted window-based self-attention, thereby enhancing efficiency and scalability. This innovation, coupled with ViTs, has revolutionized computer vision, offering state-of-the-art performance across various benchmarks and marking a paradigm shift from traditional CNN backbones to transformer-based architectures.

Specifically, several models have emerged, combining the capabilities of CNNs and transformers. TransUNet \citep{chen2021transunet} exemplifies this synergy by combining a CNN backbone for local feature extraction with a transformer module for global context understanding, supplemented by a U-Net decoder for precise segmentation. Similarly, TransClaw U-Net \citep{yao2022transclaw} integrates the strengths of vision transformers with claw-shaped convolutions and the U-Net architecture, yielding efficient and accurate segmentation results. UNETR \citep{hatamizadeh2022unetr} employs a pure transformer encoder to capture long-range data dependencies, paired with a CNN decoder for spatial precision. SwinUNetR \citep{hatamizadeh2021swin} utilizes a Swin transformer as the encoder, effectively capturing hierarchical and shift-invariant features, while a U-Net decoder reconstructs fine-grained spatial details. 
DS-TransUNet citep{lin2022ds} adopts a dual Swin transformer encoder for multi-scale feature extraction, coupled with a U-Net decoder for enhanced segmentation accuracy. HRSTNet \citep{wei2023high} further advances this approach with a Swin transformer encoder and a multi-resolution feature fusion block, alongside a U-Net decoder. These innovative models highlight the efficacy of merging CNNs and transformers in medical image segmentation, setting new benchmarks in the field.

\section{Dataset}

\subsection{Image Acquisition}
An IRB-approved retrospective review was conducted on patients with a diagnosis of acute and subacute aortic dissection (ICD-9 codes 441.01, 441.03; ICD-10 codes I71.00, I71.01) between October 2011 and March 2020, utilizing a prospectively maintained institutional database. We identified patients with uncomplicated type B aortic dissection (TBAD) based on the absence of malperfusion, rupture, rapid degeneration, or refractory pain. These TBAD patients were medically managed without surgical intervention at their first admission. Subsequent imaging was reviewed, and those without high-resolution surveillance CTA ($\leq 3mm$ slices) beyond three months from their initial hospitalization were excluded. Standard CTA comprises three phases: the non-contrast phase, the arterial phase, and the delayed phase. Initially, the non-contrast phase is taken to detect any hematoma or plaque on the vessel wall that might be concealed by iodinated contrast. This is followed by the arterial phase, where iodinated contrast medium (ICM) is rapidly injected for optimal arterial vessel visualization, ensuring the imaging coincides with peak aortic/aortic branch contrast arrival. The delayed phase, accurately timed, follows to evaluate slow-filling and venous structures. Given that many initial CTAs from TBAD patients were sourced from various local imaging centers with differing protocols, our study's criteria mandated that all three phases be captured in $\leq 3$ mm slices. 

Our dataset comprised 59 CTA images with an axial size of 512$\times$512 pixels and an isotropic in-plane resolution ranging from 0.759 mm to 1.007 mm, averaging 0.875 mm. The number of axial slices varied between 347 and 962, with a mean of 734 slices. Axial slice thickness varied from 0.8 mm to 2 mm, averaging 0.969 mm.
To expedite model training, we re-sampled the volumes to a uniform spacing of 1.5 mm$\times$1.5 mm$\times$1.5 mm.  
The ``RandCropByPosNegLabeld" function from the MONAI library \citep{monai2020monai} was used to facilitate random cropping of a fixed-sized region from a large 3D image. The cropping center can either be a foreground or background voxel, determined by a given foreground-to-background ratio. Leveraging this function, we selected random 128$\times$128$\times$28 patches from the re-sampled volumes for training, enhancing data diversity, and mitigating overfitting.

\begin{table*}[!h]                       
	\centering
	\renewcommand{\arraystretch}{1.2} 
 \scalebox{0.79}{
	\begin{tabular}{l | l}
		\toprule[0.8pt]
  \textbf{Artery} & \textbf{Arterial flow details} \\
\midrule[0.8pt]
    \multirow{3}{*}{Innominate Artery (IA)} & Branches into the right subclavian and right common carotid arteries\\
  &Right subclavian artery supplies the right posterior cerebral circulation (via right vertebral artery) and right arm \\
  &Right common carotid supplies the right cerebral cortex\\
   \hline
  Left Common Carotid Artery (LCC) & Blood supply to the left cerebral cortex\\ 
  \hline 
  Left Subclavian Artery (LSA) & Left posterior cerebral circulation (via the left vertebral artery branch) and left arm blood supply\\ 
  \hline
  Celiac Artery (CA) &  Branches into the hepatic, splenic, and left gastric arteries to supply blood to the liver, spleen, and stomach\\
  \hline
  Superior Mesenteric Artery (SMA) & The main blood supply to the intestines\\
  \hline
  Left Renal Artery (LRA) &Blood supply to the left kidney\\
  \hline
  Right Renal Artery (RRA) &Blood supply to the right kidney\\
  \hline
 Left Common Iliac Artery (LCIA)  & Branches into the left internal and external iliac arteries\\
   \hline
  Left External Iliac Artery (LEIA)& Blood supply to the left leg\\
  \hline
  Left Internal Iliac Artery (LIIA) & Blood supply to the pelvis\\
  \hline
  Right Common Iliac Artery (RCIA) & Branches into the right internal and external iliac arteries\\
  \hline
  Right External Iliac Artery (REIA) & Blood supply to the right leg\\
  \hline
  Right Internal Iliac Artery (RIIA) & Blood supply to the pelvis\\

		\bottomrule[0.8pt]
	\end{tabular}}
	\caption{Aortic branch anatomy and their blood supply.}
	\label{tab:aorta-and-its-branches}                           	
\end{table*}
\subsection{Anatomical Overview of Aortic Branches} 
\label{sect:aorta_and_branches}
The aorta is the main outflow vessel from the heart, supplying oxygenated blood to the brain, extremities, and vital internal organs essential for life. Blockages, tears, or ruptures of the aorta and its branches can lead to significant morbidity, including stroke, organ failure, and exsanguinating hemorrhage if not rapidly addressed. 
\begin{figure}[h!]%
\centering
\includegraphics[width=0.48\textwidth]{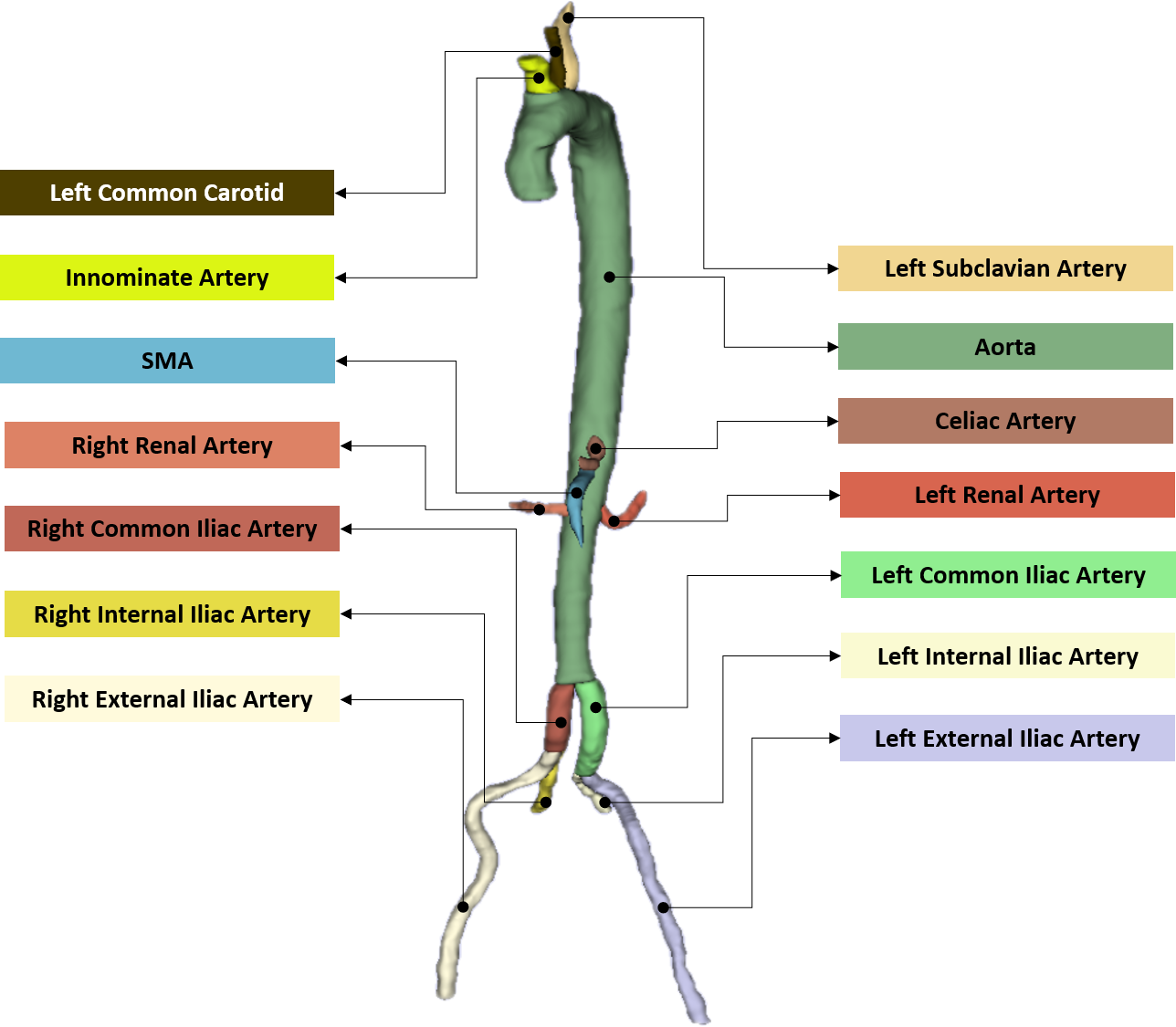}
\caption{Schematic view of the aorta and major aortic branches.}
\label{fig:aorta-branches}%
\end{figure}
Normal aortic anatomy includes thirteen primary branches, depicted in Figiure \ref{fig:aorta-branches}. The details of these branches, their respective functions, and their significance in the human body are described in Table \ref{tab:aorta-and-its-branches}.


\subsection{Data Annotation} 
\label{sect:data-annotation}
Three graduate students (VRRG, VBS, AK) worked together to manually annotate the aorta and thirteen distinct aortic branches on 3D CT images, as shown in Figiure \ref{fig:aorta-branches}, using the 3D Slicer tool \citep{fedorov20123d}. Throughout the annotation process, all three views of the CT image were utilized.
For the segmentation of elongated arteries, like the aorta and the six iliac artery branches, we primarily used the axial view, referencing the other two views as needed. First, segmentation was performed on every four axial slices. 
Subsequently, the ``fill between slices'' module in 3D Slicer was used to smoothly segment the intervening slices by interpolation.
The innominate artery, left common carotid artery, and left subclavian artery were manually segmented on all slices using the axial view. 
The celiac artery and the SMA were segmented on sagittal slices, while the left and right renal arteries were segmented on coronal slices.
Since all segmentations were performed on 2D slices, which can result in inconsistencies in 3D, we applied a 1-mm Gaussian smoothing kernel to achieve smooth segmentation in 3D.
Our initial segmentations were verified and fine-tuned by two surgery residents (JRK and BF) to obtain the ground truth annotations used for training and evaluation.
Typically, annotating a single CT scan took approximately four hours.

\begin{figure*}[h!]%
    \centering
    \includegraphics[width=\textwidth]{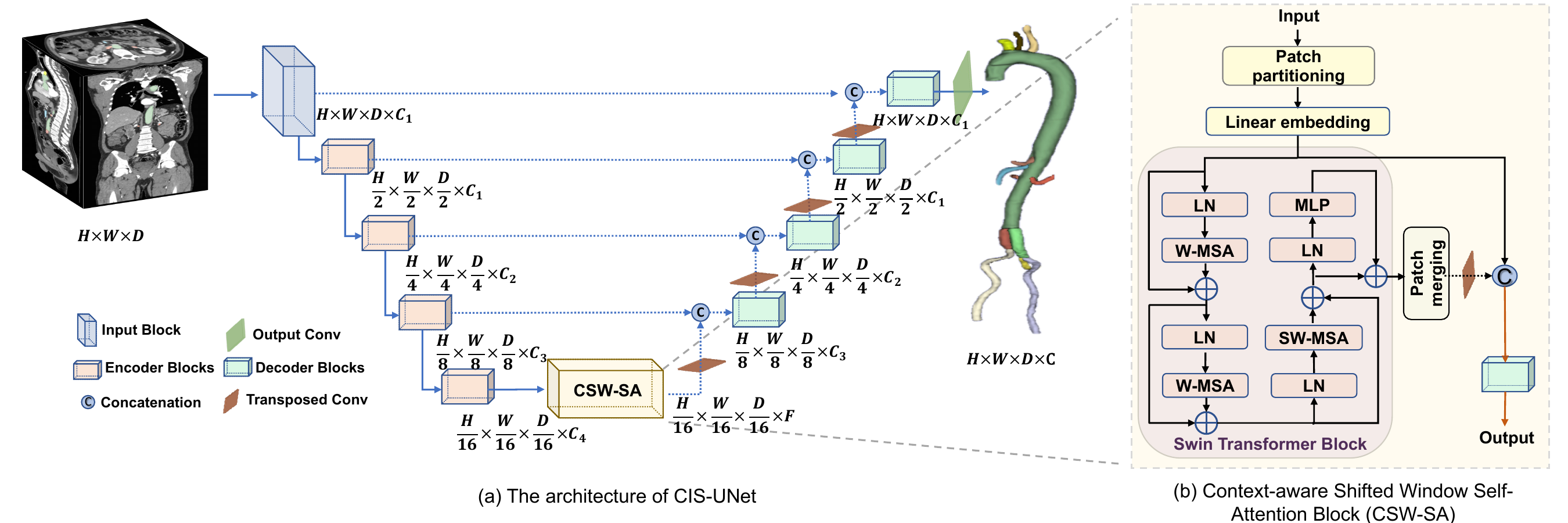}
    \vspace{-2em}
    \caption{Overview of the proposed segmentation framework. (a) architecture of the context infused Swin-UNet; (b) architecture of the context-aware shifted window self-attention block. }%
    \label{fig:CIS-UNet-and-attention}%
\end{figure*}

\section{Context Infused Swin-UNet}
In this paper, we present Context Infused Swin-UNet (\textbf{CIS-UNet}) for multi-class 3D aortic segmentation.
CIS-UNet is a hybrid segmentation network that integrates the capabilities of CNNs and the Swin transformers through a hierarchical encoder-decoder structure, ensuring computational efficiency while extracting both local and global image features.
Figure \ref{fig:CIS-UNet-and-attention} presents an overview of CIS-UNet which is comprised of a CNN encoder as well as its symmetric decoder, the skip connections, and a newly proposed Context-aware Shifted Window Self-Attention (CSW-SA) as the bottleneck block. The encoder extracts features from the input image using convolutional layers. The decoder reconstructs the segmentation map from these features via transposed convolution layers. 
Meanwhile, the self-attention block enhances feature representation by identifying long-range pixel dependencies. 
This design makes CIS-UNet a powerful architecture for aorta segmentation, which identifies and separates the elongated aorta from other structures in CT images. 

\subsection{Encoder}
\label{sect:encoder}
\begin{figure}[h!]%
\centering
\includegraphics[width=0.45\textwidth]{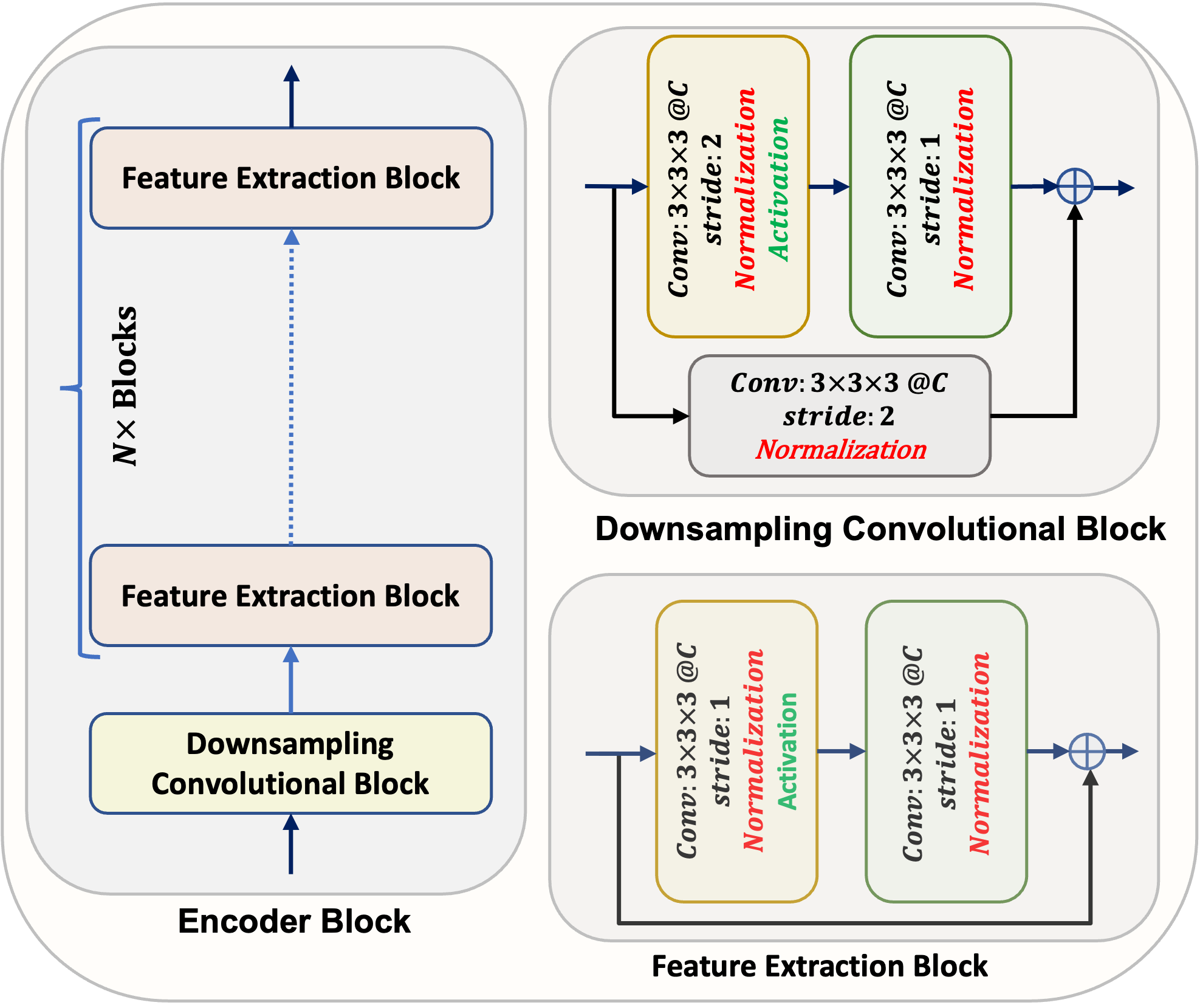}
\caption{
The encoder block.
}
\label{fig:encoderblock}
\end{figure}

The encoder of CIS-UNet comprises an input block and multiple encoder blocks. 
The input block extracts $C_1$ image features from the input image using $7\times7\times7$ convolutional kernels with a stride of 1 (see Figiure \ref{fig:CIS-UNet-and-attention}).
Each encoder block consists of a downsampling convolutional block followed by $L$ feature extraction convolutional blocks (see Figiure \ref{fig:encoderblock}). The downsampling block reduces image feature dimensions via a convolutional layer with a stride of 2 followed by another one with a stride of 1.  
The feature extraction block incorporates repeated residual convolutional units, each with two $3\times3\times3$ convolutional layers connected by a residual connection.
The number of filters in each layer of the encoder is denoted by $C_1$, $C_2$, $C_3$, and $C_4$, respectively.

\subsection{3D Context-aware Shifted Window Self-Attention}
\label{subsect:self-attention-block}
Building upon the high-level image features extracted by the CNN encoder, we introduce a novel self-attention block - Context-aware Shifted Window Self-Attention (CSW-SA) - to efficiently capture long-range dependencies between image patches.
The process begins with a pixel-wise projection of the feature map from the last layer of the encoder into a feature dimension of $F$, resulting in an input $z$ of size $\frac{H}{16} \times \frac{W}{16}\times \frac{D}{16}\times F$ to the Swin transformer block.
The Swin transformer block leverages window and shifted-window self-attention to learn the long-range image dependencies efficiently, computed as:
\begin{align}
    &\hat{z} = \text{W-MSA}(LN(z)) + z \nonumber, \\
    & z' = \text{MLP}(\text{LN}(\hat{z})) + \hat{z} \nonumber,\\
    &\bar{z} = \text{SW-MSA}(\text{LN}(z') + z', \nonumber\\
    &z'' = \text{MLP}(\text{LN}(\bar{z})) + \bar{z},  \nonumber
\end{align}
where W-MSA and SW-MSA denote window-based multi-head self-attention and shifted-window multi-head self-attention, MLP denotes multilayer perceptron, and LN denotes layer normalization.

However, the use of local window-based self-attention imposes limitations on the model's capacity to effectively capture global dependencies. To improve the model's ability to extract and integrate global contextual information across self-attention windows, our proposed CSW-SA is designed to enhance the original window-based self-attention by incorporating global context through repurposed patch merging.

\begin{figure}[h!]%
\centering
\includegraphics[width=0.48\textwidth]{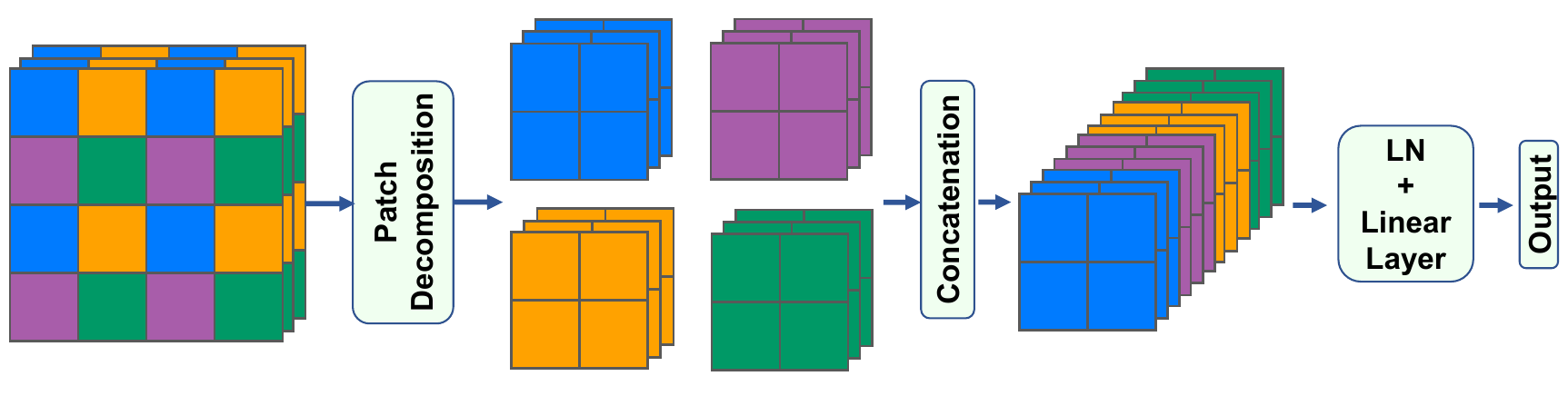}
\caption{Illustration of the patch merging layer.}
\label{fig:patch-merging-illustration}%
\end{figure}

The patch merging layer, a key component in the original Swin Transformer architecture, plays a crucial role in spatial dimension reduction and channel augmentation. As depicted in Figiure \ref{fig:patch-merging-illustration}, this operation involves merging neighboring patches into larger ones and concatenating them along the channel dimension.
Specifically, here in our CSW-SA, unlike the usage in conventional swin Transformers, as shown in Figiure \ref{fig:CIS-UNet-and-attention}(b), patch merging is employed to condense the map $z''$ into dimensions of $\frac{H}{32} \times \frac{W}{32} \times \frac{D}{32} \times 2F$, effectively providing a global spatial context. Subsequently, a transposed convolution layer is introduced, upscaling the condensed feature map to its original size of $\frac{H}{16} \times \frac{W}{16} \times \frac{D}{16} \times F$. This upsampled map is then merged with the output from the linear embedding (i.e., $z$).
The final representation undergoes refinement through two consecutive $3\times3\times3$ convolutional kernels, resulting in an output feature map of dimensions $\frac{H}{16} \times \frac{W}{16} \times \frac{D}{16}\times F$, as illustrated in Figiure \ref{fig:CIS-UNet-and-attention}. This refined output seamlessly integrates into the decoder of the CIS-UNet for subsequent upsampling.

Unlike the utilization of shift-based self-attention in the Swin-UNet and Swin-UNETR models at every downsampling stage, our CSW-SA block employs just one Swin transformer block at the bottleneck layer for enhanced computational efficiency. Our experimental results will demonstrate how this design enhances performance without incurring unnecessary computational costs.
    
\subsection{Decoder}
\label{subsect:decoder_path}
The decoder is responsible for reconstructing a high-resolution segmentation map from the low-resolution feature maps. It consists of four transposed convolution layers and four decoder blocks. The transposed convolution layers progressively upsample the input by a factor of $2$, enlarging the spatial resolution and allowing for finer segmentation details. 
The decoder blocks perform a series of operations: feature concatenation, standard convolution, and residual connection. The feature concatenation integrates features from the corresponding encoder layer via a skip connection. 
The standard convolution consists of two $3\times 3\times 3$ convolutions with a stride of $1$. 
The final convolution layer has filters of size $1\times1\times1$, which produces the final probability segmentation map of size $H\times W\times D \times C$, where $C$ is the number of classes to be segmented.

\subsubsection{Loss Function}
\label{sect:loss-function}
To train our model, we use a loss function that combines the Dice loss and the cross-entropy loss. The Dice loss measures the relative overlap between the ground truth and the predicted masks, while the cross entropy loss measures the pixel-wise uncertainty. 
We use the DiceCELoss function from MONAI \citep{monai2020monai}, a PyTorch-based framework for medical image analysis, which is defined as follows:
\begin{equation}
    \label{eq:dice-ce-loss}
    L_{DCE} = \lambda_{Dice}L_{Dice} + \lambda_{CE} L_{CE},
\end{equation}
where $\lambda_{Dice}$ and $\lambda_{CE}$ are the weights for the Dice loss and the cross-entropy loss, respectively. We assigned equal weights to both the Dice loss and the cross-entropy loss. This weighting scheme establishes a balance between segmentation accuracy and pixel-wise classification.

The Dice loss \citep{ma2021loss} is computed as:
\begin{equation*}
    \label{eq:dice-loss}
    L_{Dice} = 1 - \frac{2\sum_{c=1}^{C}\sum_{i=1}^{N}g_i^cs_i^c}{\sum_{c=1}^{C}\sum_{i=1}^{N}g_i^c + \sum_{c=1}^{C}\sum_{i=1}^{N}s_i^c},
\end{equation*}
where $C$ is the number of classes, $N$ is the number of voxels, $g_i^c$ is the binary value (0 or 1) of the voxel at index $i$ of class label $c$ in the ground truth multi-label mask, and $s_i^c$ is the probability value (between 0 and 1) of the voxel at index $i$ of class label $c$ in the predicted multi-label mask. 

The cross-entropy loss \citep{ma2021loss} is computed as:
\begin{equation*}
    \label{eq:ce-loss}
    L_{CE} =  -\frac{1}{N}\sum_{c=1}^{C}\sum_{i=1}^{N}g_i^c\log s_i^c,
\end{equation*}
where the variables are the same as in the Dice loss equation.

\subsection{Implementation Details}
\label{sect:implementation-details}
We split our dataset of 59 CT image volumes randomly into two subsets: 44 volumes for training and 15 for testing.  Our training pipeline was implemented using the PyTorch framework and the MONAI library. 
To effectively handle these large volumes, we utilized the `sliding window inference' technique from MONAI, with a patch size of $128\times128\times128$ and a batch size of 4.
For optimization, we chose the AdamW optimizer \citep{loshchilov2017decoupled} with a learning rate of $10^{-4}$ and a weight decay of $10^{-5}$.  Our model was trained for 3000 iterations, roughly equivalent to 666 epochs. 
For the encoder, we chose $C_1 = 64$, $C_2 = 128$, $C_3 = 256$, $C_4 = 512$.
The feature embedding dimension in the self-attention block was chosen as $F = 48$.
Training was performed on a single NVIDIA A100 GPU with 80 GB RAM.

\subsection{Model Evaluation}
In line with previous aorta segmentation research \citep{fantazzini20203d,chen2021multi}, we assessed our model using the Mean Surface Distance (MSD) and Dice Similarity Coefficient (DSC) for each branch, averaging the results across all subjects.
MSD calculates the average distance between the surfaces of the original and predicted segmentation maps as follows:
\begin{equation}
    \label{eq:msd}
    MSD = \frac{1}{N}\sum_{p\in Y}\left(\min_{q\in \hat{Y}} d(p, q)\right),
\end{equation}
where $Y$ and $\hat{Y}$ are the original and predicted segmentation surfaces, $N$ is the number of points on $Y$, and $d(p,q)$ is the Euclidean distance between the points $p$ and $q$. 

The DSC measures the relative overlap between two segments.
It ranges from $0$ (no overlap) to $1$ (perfect overlap), and can be computed as follows:
\begin{equation}
    \label{eq:dice_coefficient}
    DSC(Y, \hat{Y}) = \frac{2|Y \cap \hat{Y}|}{|Y|+|\hat{Y}|},
\end{equation}
where $Y$ and $\hat{Y}$ are the original and the predicted segmentation masks, and $|\cdot|$ denotes the cardinality of a set. 

\section{Experiments and Results}
\label{sect:results}

We evaluated our CIS-UNet model against several leading 3D segmentation models including 3D-UNet \citep{kerfoot2019left}, SwinUNetR \citep{tang2022self}, dResNet \citep{raza2023dresu}, and UNetR \citep{hatamizadeh2022unetr}. 

\subsection{Quantitative Results}
Table \ref{tab:dice} demonstrates that CIS-UNet achieved the highest average DSC of $0.713$ across 14 branches, surpassing SwinUNetR's DSC of $0.697$. Notably, CIS-UNet was the leading model in 9 out of these 14 branches. In instances where CIS-UNet's DSC marginally trailed behind SwinUNetR, the differences were negligible, with variances of only $0.031$ for LCC, $0.001$ for REIA, and $0.003$ for RIIA. 
\begin{table}[!h]                       
	\centering
	\renewcommand{\arraystretch}{1.3} 
 \scalebox{0.85}{
	\begin{tabular}{c | c c c c c}
		\toprule[0.8pt]
  \multirow{2}{*}{\rotatebox[origin=c]{45}{\textbf{Branches}}} &  \multicolumn{5}{|c}{Methods}  \\
  \cmidrule(lr){2-6}  
   &  3D-UNet & SwinUNetR & dResNet & UNetR & \textbf{CIS-UNet} \\
		\midrule[0.8pt]
		Aorta     &0.908      &0.913                &0.920      &0.897    & \textbf{0.922}           \\
        IA        &0.729      &0.728                &0.728       &0.681    &\textbf{0.741}     \\
        LCC       &0.635      &\textbf{0.657}       &0.612      & 0.582    & 0.644              \\
        LSA       &0.75       &0.753                &0.776      & 0.782    &\textbf{0.792}      \\
        CA        &0.57       &\textbf{0.622}       &0.567      & 0.569    & 0.58               \\
        SMA       &0.723      &0.691                &\textbf{0.766} & 0.669 & 0.715              \\
        LRA       &0.503      &0.527                &0.47         & 0.423  & \textbf{0.54}      \\
        RRA       &0.507      &0.584                &0.534        & 0.586  & \textbf{0.594}     \\
        LCIA      &0.786      &0.807                &0.788       & 0.766    & \textbf{0.837}    \\
        RCIA      &0.74       &0.740                &0.737        & 0.653  &\textbf{ 0.788}     \\
        LEIA      &0.743      &0.776                &0.783        & 0.745  &\textbf{0.805}       \\
        REIA      &0.702      &\textbf{0.784}       &0.774       & 0.726   &0.783               \\
        LIIA      &0.586      &0.606                &0.625        & 0.585  &\textbf{0.666}       \\
        RIIA      &0.514      &\textbf{0.573}       &0.546        & 0.492  &0.57                \\ 
        \midrule[0.8pt]
        Average    &0.671&0.697&0.688 &0.654 &\textbf{0.713}    \\
		\bottomrule[0.8pt]
	\end{tabular}}
	\caption{Dice similarity coefficients (DSC) for various 3D aortic segmentation models.}
	\label{tab:dice}                           	
\end{table}

The mean surface distance (MSD) for each of the 14 aortic branches is detailed in Table \ref{tab:msd}. Our CIS-UNet model notably achieved the lowest MSD in 9 of these 14 branches. With an average MSD of 2.767 mm, it surpassed SwinUNetR, the second-best performer, which had an MSD of 3.3394 mm, by 18.47\%. While acknowledging areas for improvement in our model, particularly in regions like SMA, REIA, and RIIA that present segmentation challenges due to their small size, complex shape, and anatomical variability.
\begin{table}[!h]                       
	\centering
	\renewcommand{\arraystretch}{1.3} 
  \scalebox{0.85}{
	\begin{tabular}{c | c c c c c}
		\toprule[0.8pt]
  \multirow{2}{*}{\rotatebox[origin=c]{45}{\textbf{Branches}}} &  \multicolumn{5}{|c}{Methods}  \\
  \cmidrule(lr){2-6}  
   &  3D-UNet & SwinUNetR & dResNet & UNetR & \textbf{CIS-UNet} \\
		\midrule[0.8pt]
		Aorta       &0.816      &1.032          &0.711        &1.292    &\textbf{0.666}             \\
        IA          &1.54       &3.81           &1.201         &1.319 &\textbf{1.184}                 \\
        LCC         &1.662      &2.048          &2.272        &1.995   &\textbf{1.454}          \\
        LSA         &$\dagger$  &1.071          &0.785         &\textbf{0.682} &0.698                      \\
        CA          &2.189      &2.415          &$\dagger$     &2.107 &\textbf{1.97}            \\
        SMA         &2.212      &4.987          &\textbf{1.003}  &5.294 &1.497                        \\
        LRA         &2.85       &5.26           &$\dagger$    &4.369  &\textbf{2.703}            \\
        RRA         &$\dagger$  &10.312         &$\dagger$    &4.615   &\textbf{1.759}     \\
        LCIA        &1.352      &1.609          &0.997        &3.78   &\textbf{0.773}             \\
        RCIA        &2.294      &2.743          &\textbf{1.654}&6.248 &1.966                     \\
        LEIA        &4.58       &1.84           &2.881         &3.42  &\textbf{1.413}              \\
        REIA        &7.104      &\textbf{2.97}  &1.706         &3.328 &13.46                         \\
        LIIA        &3.066      &4.099          &2.467         &6.297  &\textbf{1.478}             \\
        RIIA        &7.156      &\textbf{3.321} &$\dagger$      &8.14  &7.711                \\        
        \midrule[0.8pt]
        Average &4.058      &3.394      &3.976     &3.778 &\textbf{2.767}              \\
		\bottomrule[0.8pt]
	\end{tabular}}\\
        \footnotesize{$\dagger$  represent outliers for the branches where the model failed to segment.}\\
	\caption{Average mean surface distances (mm) of different 3D aortic segmentation models.}
	\label{tab:msd}                           	
\end{table}

The superior Dice coefficient and reduced average surface distance achieved by CIS-UNet suggest not only a higher global overlap but also enhanced boundary prediction accuracy. Therefore, we conclude that our proposed architecture outperforms other existing models in segmentation quality and robustness.

\subsection{Qualitative Results}

\begin{figure*}[h]%
\centering
\includegraphics[width=0.95\textwidth]{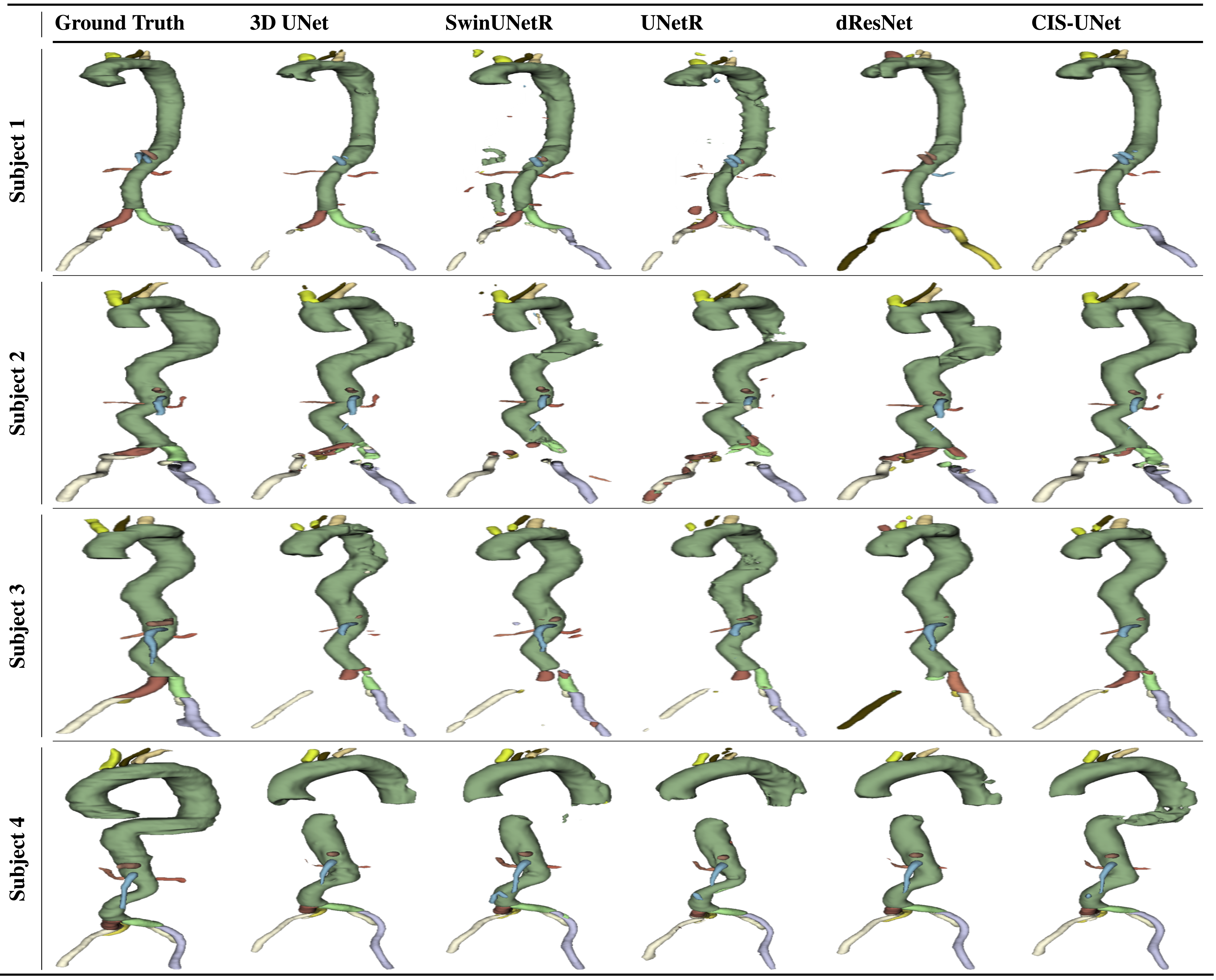}
\caption{Aorta segmentation results on four complex cases.  Our CIS-UNet model achieved superior performance by accurately and consistently segmenting most of the branches. Other models had limitations, such as failing to segment important branches or producing discontinuous segmentation.}
\label{fig:difficult-qualitative-results}%
\end{figure*}

Figure \ref{fig:difficult-qualitative-results} illustrates the segmentation outcomes for four complex cases with unusual anatomy or thoracoabdominal aortic aneurysms, which are less common in clinical practice and underrepresented in our dataset. It is evident that current segmentation methods struggled, showing issues like misclassified branches, discontinuities in segments, or noticeable artifacts. Crucially, Figure \ref{fig:difficult-qualitative-results} highlights the superior performance of our proposed model in accurately segmenting the aorta and its branches in these particularly challenging scenarios. Below, we provide a detailed analysis of the segmentation result for each case.
\begin{itemize}
    \item \textbf{Subject 1:} This subject has a thoracoabdominal aneurysm, or a pathological dilation of the aorta between the subclavian and celiac arteries. Several models produced segmentation artifacts, missed important branches, or resulted in discontinuous segmentation of the aorta. For example, 3D UNet \citep{kerfoot2019left} was unable to segment the celiac artery and produced discontinuities in the segments of the right and left external iliac arteries. SwinUNETR's \citep{hatamizadeh2021swin} segmentation includes several notable artifacts and also produces a break in the right external iliac artery. UNETR \citep{hatamizadeh2022unetr} misclassified the celiac artery, failed to segment a wedge of the proximal thoracic aorta, and also had several obvious artifacts. dResNet \citep{raza2023dresu} misclassified the right external iliac artery, the left common carotid artery, and several other branches. While our model misclassified the celiac artery and had a break in the left renal artery, all other branches were correctly identified with no artifacts or large breaks in continuous segments.
    \item \textbf{Subject 2:} This subject has a significant amount of baseline aortic tortuosity and a large descending thoracic aortic aneurysm, which can often accompany acute aortic dissection. Most models failed to correctly segment this case or produced discontinuous segmentation. For example, 3D UNet, SwinUNETR, UNETR, and dResNet all incompletely segmented the thoracic aneurysm and produced breaks in the iliac arteries. Only our model successfully segmented the tortuous and aneurysmal portion of the aorta as a continuous structure.
    \item \textbf{Subject 3:} This subject has a tortuous abdominal aorta, and as a result, the origins of the celiac artery, superior mesenteric artery, and left renal artery take off at atypical angles from the abdominal aorta. Most models could not correctly identify these branches. For example, 3D UNet, in addition to an incomplete aorta segmentation, captured only a small portion of the celiac artery and missed a large portion of the right common iliac artery. SwinUNETR mislabeled the left common carotid artery and had prominent gaps in the segmentation of the right and left common iliac arteries. UNETR had large gaps in several aortic branches. dResNet segmented the celiac artery as part of the liver and failed to segment several small branches. While our model did not segment the left renal artery, it correctly identified all other branches with appropriate boundaries and minimal gaps.
    \item \textbf{Subject 4:} This subject has a large thoracoabdominal aortic aneurysm, a tortuous aorta, and the celiac, superior mesenteric, right renal, and left renal arteries arise at atypical locations from the abdominal aorta. All existing models failed to capture the aortic tortuosity, and most models could not segment all four branches correctly or produced inaccurate segmentation. For example, while 3D UNet and SwinUNETR identified the celiac, superior mesenteric, right renal, and left renal branches, the boundaries were incompletely detected, and SwinUNETR inappropriately segmented a duplicate branch of the superior mesenteric artery. UNETR and dResNet failed to identify one or more of the celiac, superior mesenteric, right renal, and left renal artery branches. While our model also inappropriately segmented a second small portion of the superior mesenteric artery, only our proposed model could identify and segment all these branches with accurate boundaries while also capturing the extreme tortuosity of the aorta.
\end{itemize}

\subsection{Comparative Analysis of Model Efficiency}
\label{subsect:computational-complexity}
We assessed the efficiency of 3D segmentation models by examining their number of trainable parameters and the time taken to infer segmentation results for each 3D CT image volume. As detailed in Table \ref{tab:computational-complexity}, all models have a comparable number of parameters, with dResNet having the highest at 94.38 million and SwinUNetR having the lowest at 61.99 million. In terms of speed, UNet and dResNet are notably efficient, processing at 13 ms and 20 ms per 3D volume, respectively. Despite having only 61.99 million parameters, SwinUNetR requires the longest time, at 125 ms per run, primarily due to the computationally intensive self-attention layers in its encoder layers. CIS-UNet strikes a balance between efficiency and accuracy, possessing the second-smallest number of trainable parameters and the third-fastest inference time. This balance is achieved by incorporating the Swin transformer in the bottleneck layer rather than in the encoder/decoder layers, where the input image sizes are significantly larger.

\begin{table}[!h]
\centering
\renewcommand{\arraystretch}{1.3} 
\scalebox{0.85}{
\begin{tabular}{c| c | c | c  }
\toprule[0.8pt]
Models      & Avg. DSC & \# Params (M)     & Inference Time (ms) \\
\midrule[0.8pt]
SwinUNetR   & 0.697   &61.99      & 125\\
3D-UNet      &0.671   &77.16      & 13\\
UNetR        &0.654   &92.618     & 49\\ 
dResNet      &0.688   &94.375     & 20\\
CIS-UNet (Ours)  &0.713  &75.038     & 63\\
\bottomrule[0.8pt]
\end{tabular}}
\caption{Comparative analysis of different 3D image segmentation models in terms of parameter efficiency and inference speed.}
\label{tab:computational-complexity}                           	
\end{table}

\subsection{Ablation Study}
\label{sect:ablation-study}
The proposed CIS-UNet model has three hyperparameters to control its efficiency and accuracy: the number of feature extraction blocks in each layer of the encoder ($L$), the number of convolutional filters in each of the encoder and decoder layers ($C$), the feature embedding size of the Swin transformer ($F$). We evaluate the performance of several model variants with distinct sizes, namely Tiny, Small, and Base. 
To demonstrate the effectiveness of our proposed CSW-SA, we conduct a comparative analysis between the Base model utilizing the original shifted window self-attention (SW-SA) in Swin Transformers and our proposed CSW-SA. 
The specific architecture hyperparameters of these model variants are detailed as follows:
\begin{itemize}[itemindent=1pt]
   \item[-] Tiny: L = (2,2,2,2), C = (32,64,128,56), F = 48
    \item[-] Small: L = (3,4,6,3), C = (32,64,128,56), F = 48
        \item[-] Base: L = (3,4,6,3), C = (64,128,256,512), F = 48
\end{itemize}

\begin{figure}[h!]%
\centering
\includegraphics[width=0.45\textwidth]{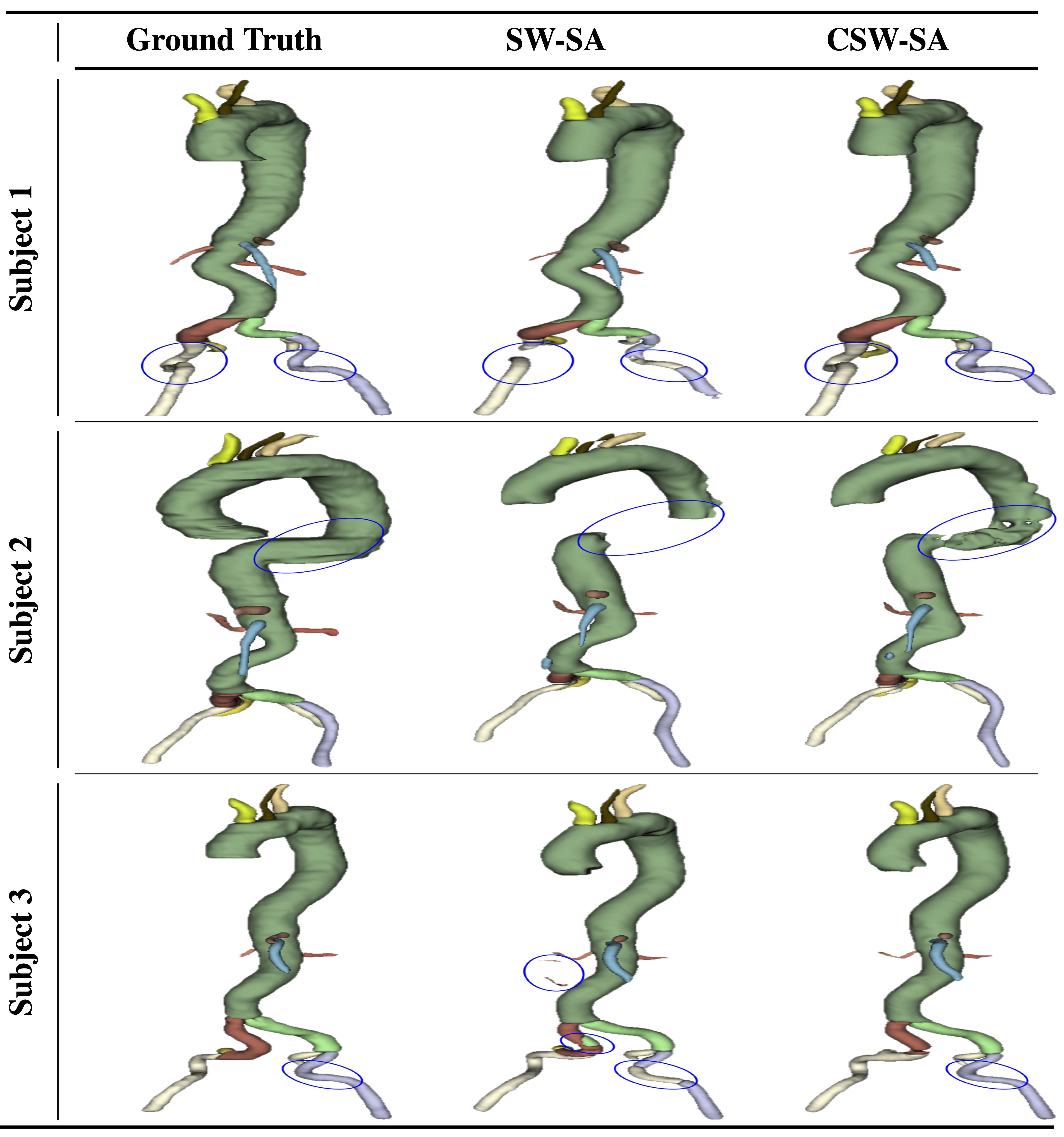}
\caption{Qualitative comparison of segmentation results using CSW-SA and SW-SA.}
\label{fig:patch-merging-effects}%
\end{figure}

\begin{table}[!h]                       
	\centering
	\renewcommand{\arraystretch}{1.3} 
 \scalebox{0.85}{
	\begin{tabular}{c|c|c | c | c  }
		\toprule[0.8pt]
        CIS-UNet &CSW-SA & SW-SA & Avg. DSC  & \#Params (M) \\
		\midrule[0.8pt]
            Tiny  & \checkmark & & 0.694 & 13.921    \\
            Small &\checkmark & &0.697  & 21.5     \\
            Base  & &\checkmark& 0.701 & 71.789   \\
            Base  &\checkmark & & 0.713 & 75.038   \\
         \bottomrule[0.8pt]
	\end{tabular}}
	\caption{\scriptsize{Comparison of the performance of different model variants.}}
	\label{tab:model-comparison}                    	
\end{table}
Table \ref{tab:model-comparison} shows the Dice coefficient (DSC) and the number of parameters for each model variant. The results suggest that improving the model size by increasing the number of convolutional filters and feature extraction blocks enhances the segmentation accuracy. We recognize that these improvements come at the cost of increased model complexity and computational time. The comparison between CSW-SA and SW-SA highlights the importance of exploiting global context in the window-based self-attention. This strategy significantly improves the Dice coefficient from 0.701 to 0.713, with a small increase in model complexity from 71 million parameters to 75 million parameters. 
In Figure \ref{fig:patch-merging-effects}, a qualitative comparison of segmentation results for the base model using CSW-SA and SW-SA is presented. The benefits of our proposed CSW-SA are clear, as it prevents broken segmentation, artifacts, and misclassification, leading to smoother, more continuous, and accurate segmentation.
To summarize, based on Tables~\ref{tab:computational-complexity} and \ref{tab:model-comparison}, we can conclude that CSW-SA stands out as the most advantageous choice for aortic segmentation compared with existing Transformers' self-attention approaches, as it achieves the best balance between segmentation accuracy, model complexity, and computational efficiency.

\subsection{Generalization to Other Segmentation Problems}
\label{sect:generalization}
\begin{figure}[h!]%
    \centering
    \subfloat[\centering Ground Truth \label{fig:OriginalBTCV}]{{\includegraphics[width=0.225\textwidth]{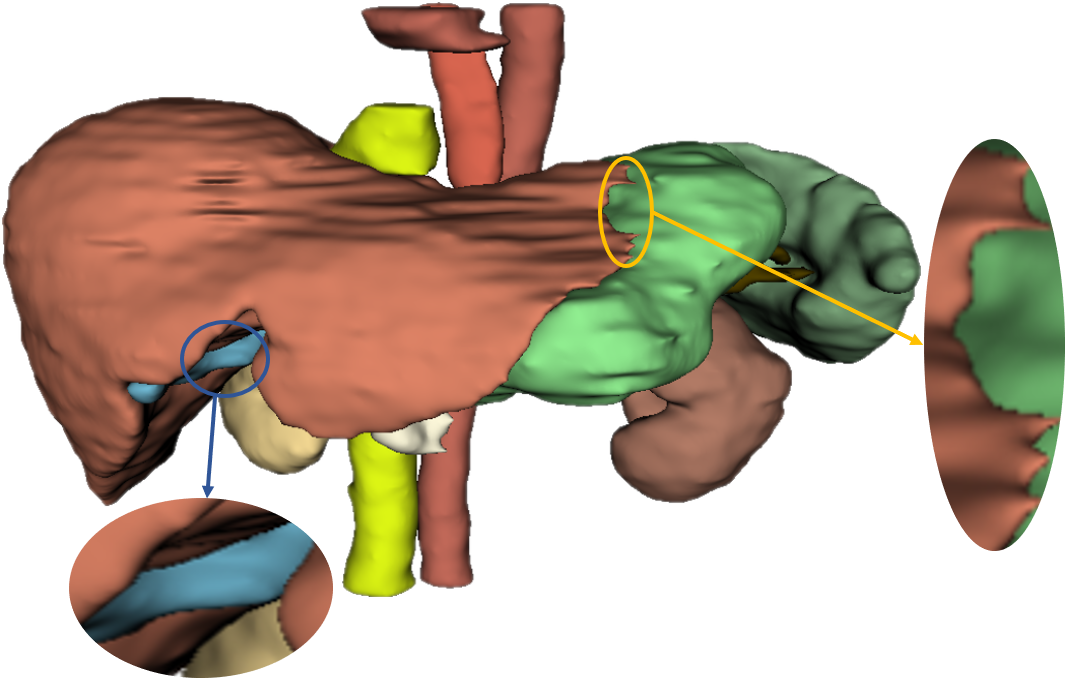}}}%
    \quad
    \subfloat[\centering UNetR Segmentation \label{fig:UNetRBTCV}]{{\includegraphics[width=0.225\textwidth]{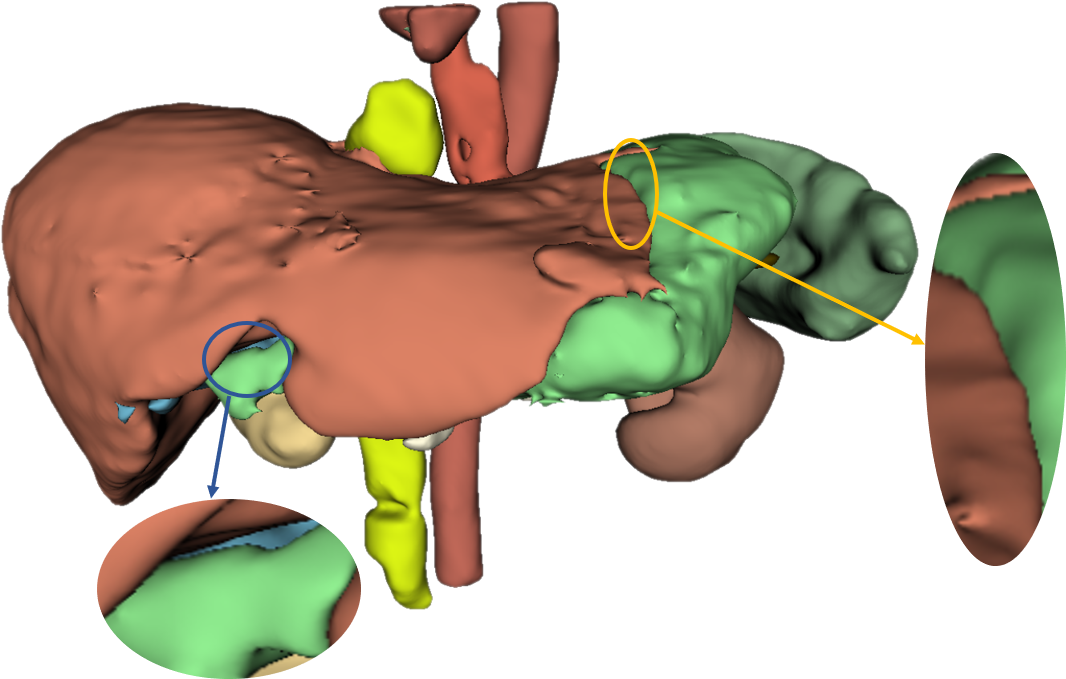}}}%
    \quad
    \subfloat[\centering SwinUNetR Output \label{fig:SwinUNetRBTCV}]{{\includegraphics[width=0.225\textwidth]{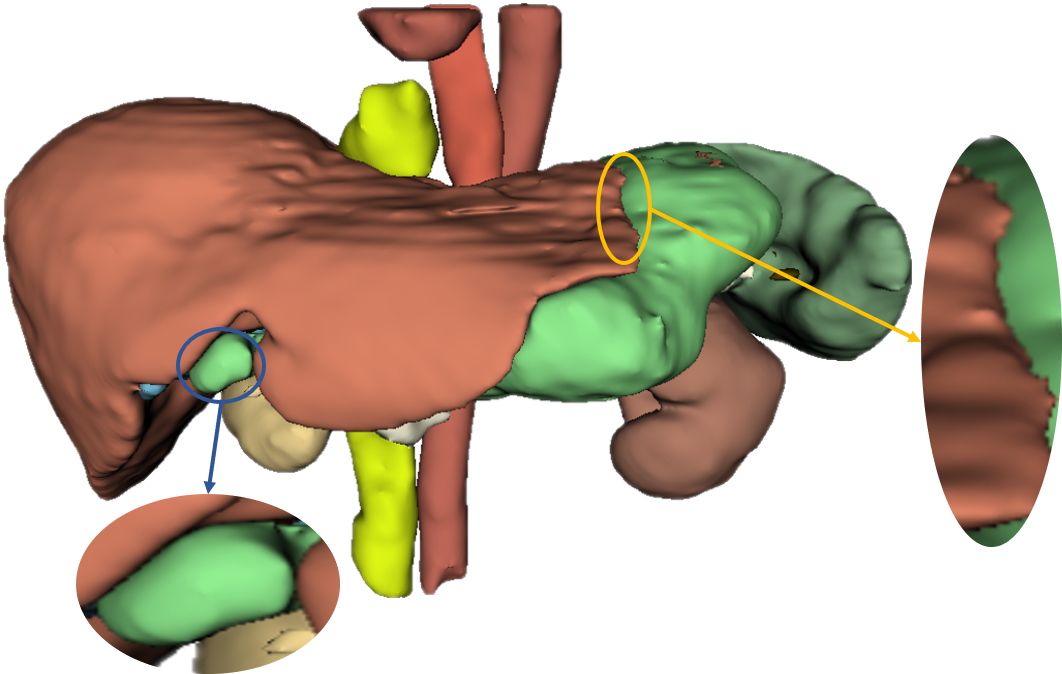} }}%
    \quad
    \subfloat[\centering CIS-UNet Output \label{fig:OurBTCV}]{{\includegraphics[width=0.225\textwidth]{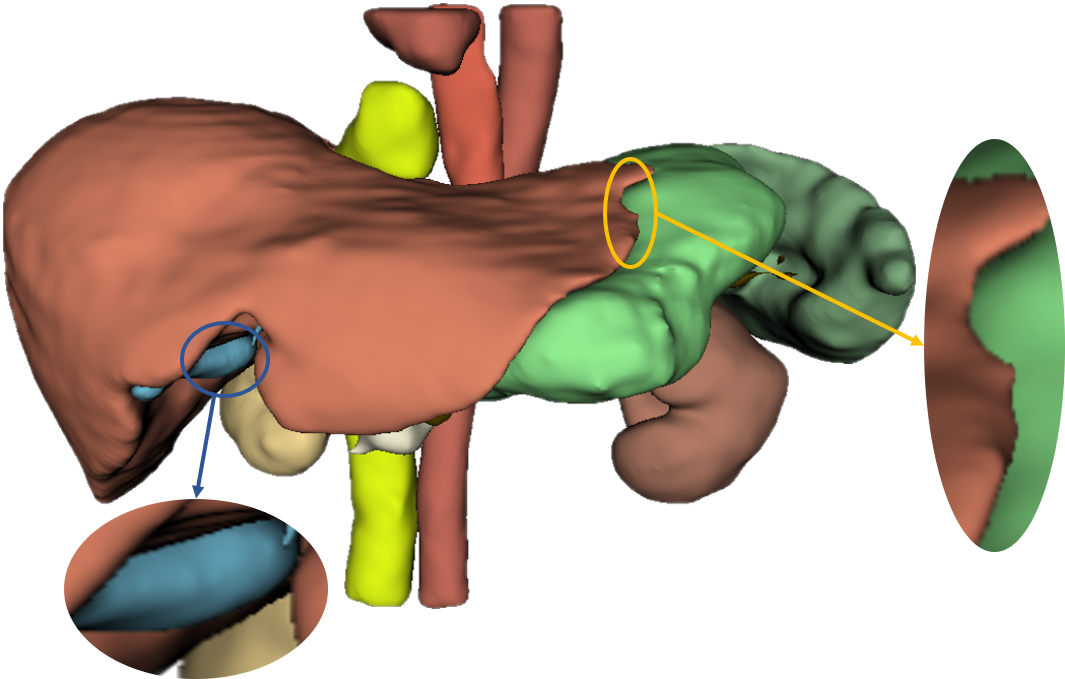} }}%
    \caption{Visual comparison of segmentation results on a representative case from the BTCV dataset. The circles indicate regions where UNetR and SwinUNetR misclassified or over-segmented organs.}%
    \label{fig:visual-analysis_btcv}%
\end{figure}

To assess the generalizability of our CIS-UNet model to similar 3D segmentation problems, we evaluated its performance against two transformer-based segmentation models, SwinUNetR and UNetR, using the publicly accessible BTCV dataset \citep{landman2015miccai}. The BTCV dataset comprises scans from 30 subjects who underwent abdominal CT. Clinical radiologists at Vanderbilt University Medical Center annotated 13 organs in these scans: spleen (spl), right kidney (rkid), left kidney (lkid), gallbladder (gall), esophagus (eso), liver (liv), stomach (sto), aorta (aor), inferior vena cava (IVC), portal and splenic veins (veins), pancreas (pan), right adrenal gland (RAG), and left adrenal gland (LAG). All models were trained for 1000 epochs, with 25 subjects for training and 5 for validation.
\begin{table*}[!h]                       
	\centering
	\renewcommand{\arraystretch}{1.3} 
 \scalebox{0.82}{
	\begin{tabular}{c | c c c c c c c c c c c c c c | c }
		\toprule[0.8pt]
            Models & spl & rkid & lkid & gall & eso & liv & sto & aor & IVC & veins & pan & RAG  & LAG & \textbf{Avg} & Metric\\
		\midrule[0.8pt]
            UNetR  & 0.908 & 0.926 & 0.927 & 0.68 & 0.712 & 0.951 & 0.72 & 0.871 & 0.792 & 0.667 & 0.745 & 0.638 & 0.604 & 0.78 & \multirow{3}{*}{\rotatebox[origin=c]{90}{\textbf{DSC}}} \\ 
            SwinUNetR  & 0.957 & 0.947 & 0.937 & 0.67 & 0.761 & 0.968 & 0.819 & 0.9 & 0.85 & 0.742 & 0.807 & 0.7 & 0.583 & 0.819 & \\
            CIS-UNet & \textbf{0.958} & 0.945 & \textbf{0.943} & \textbf{0.734} & 0.76 & 0.968 & \textbf{0.869} & 0.899 & \textbf{0.864} & 0.74 & \textbf{0.811} & 0.682 & \textbf{0.687} & \textbf{0.835} & \\
        \midrule[0.8pt]
            UNetR  &  4.111 & 1.99 & 1.31 & 1.138 & 2.819 & 5.265 & 4.734 & 1.248 & 5.285 & 1.482 & 2.267 & 0.778 & 2.807 & 2.71 & \multirow{3}{*}{\rotatebox[origin=c]{90}{\textbf{MSD (mm)}}} \\
            SwinUNetR  & 0.472  & 0.452 & 2.018 & 1.939 & 1.128 & 0.869 & 2.625 & 0.676 & 0.802 & 1.048 & 0.968 & 0.486 & 1.242 & 1.13 & \\
            CIS-UNet & \textbf{0.428} & \textbf{0.389} & \textbf{0.396} & 1.588 & \textbf{1.007} & \textbf{0.862}&\textbf{1.819} & \textbf{0.649} & \textbf{0.715} & \textbf{0.988} & \textbf{0.925} & 1.544 & \textbf{0.778} & \textbf{0.93} & \\  
		\bottomrule[0.8pt]
	\end{tabular}}
	\caption{\scriptsize{Comparison of CIS-UNet, SwinUNetR, and UNetR in multi-organ segmentation on the BTCV dataset.}}
	\label{tab:monai-results}                           	
\end{table*}
Results in Table \ref{tab:monai-results} demonstrate that our CIS-UNet model outperformed both SwinUNetR and UNetR across both the Dice coefficient and the mean surface distance (MSD). Figure \ref{fig:visual-analysis_btcv} presents the segmentation results of a representative subject. This figure indicates that while all models successfully segmented the BTCV dataset, there are some notable differences. The UNetR and SwinUNetR models misclassified and over-segmented two regions, marked by the circles, whereas our model effectively avoided such errors.
We acknowledge the differences between the BTCV dataset and our aorta dataset. In our aorta dataset, there are smaller branches such as the RIIA, LIIA, LRA, and RRA, whereas most of the 13 organs in the BTCV dataset are significantly larger. Nevertheless, the consistently superior performance of our CIS-UNet on both datasets underscores its potential in various 3D multi-class segmentation tasks.

\section{Discussion}
\subsection{Clinical Implications}
Accurate modeling of the aorta and its branch vessels is critical for minimally invasive treatment of aortic disease.  This is particularly true as we are treating increasingly complex aortic pathologies that extend into the aortic arch and/or perivisceral segment and require branched/fenestrated endografts to exclude flow from the aneurysm but maintain branch vessel flow.  Inaccurate modeling can lead to an inability to appropriately plan treatment and result in branch vessel coverage with highly morbid consequences.  While semi-automatic tools exist for detailed planning that incorporate branch vessels, a fully automated system would allow for more efficient and accurate planning.  Furthermore, the ability to use publicly available software instead of a commercial product will increase access to complex aortic planning resources, potentially expanding access to care in this vulnerable patient population. 

The architecture detailed in this paper is fully automated, surpassing other similar programs in precision and consistency. Thus, this architecture could form the foundation for a tool assisting preoperative planning for patients needing endovascular treatment for aortic pathologies, ensuring accurate mapping for all, simplifying training for surgeons, and streamlining surgical preparations. 

\subsection{Limitations of Our Study}
We recognize two major limitations in our study. First, all of our CT images originate from the aortic dissection cohort, potentially impacting the generalizability of our model to other aortic diseases, such as those with primary aneurysmal disease or those with prior surgical repairs. To address this in the future, we aim to employ a larger and more diverse dataset. Second, the manual annotation process is both time-consuming and labor-intensive, averaging 4 hours for each CT image volume. To streamline this process, we intend to use our CIS-UNet model for the preliminary automatic annotation of new CT images. While these initial annotations might not achieve perfection, they can significantly reduce the subsequent manual correction efforts.

\section{Conclusion}
In this paper, we introduce CIS-UNet, a novel architecture for 3D aortic segmentation that amalgamates the strengths of convolutional neural networks and vision transformers. Tested on a dataset of 59 subjects with aortic dissection, our model outperformed four state-of-the-art segmentation models. Ablation studies revealed significant performance enhancements with the proposed context-aware shifted window self-attention. Additionally, CIS-UNet demonstrated robust generalization to another 3D segmentation dataset, surpassing two recent vision transformer-based segmentation models. These findings underscore CIS-UNet's superior performance and its potential for diverse image segmentation applications.

\section*{Authorship Credit}
\textbf{Muhammad Imran:} Data Curation, Formal analysis, Investigation, Methodology, Software, Validation, Visualization, Writing - Original Draft
\textbf{Jonathan R Krebs:} Data Curation, Writing - Original Draft
\textbf{Veera Rajasekhar Reddy Gopu:} Data Curation, Writing - Review \& Editing
\textbf{Brian Fazzone:} Data Curation, Writing - Review \& Editing
\textbf{Vishal Balaji Sivaraman:} Data Curation, Writing - Review \& Editing
\textbf{Amarjeet Kumar:} Data Curation, Writing - Review \& Editing
\textbf{Chelsea Viscardi:} Data Curation, Writing - Review \& Editing
\textbf{Robert Evans Heithaus:} Data Curation, Writing - Review \& Editing
\textbf{Benjamin Shickel:} Supervision, Writing - Review \& Editing
\textbf{Yuyin Zhou:} Methodology, Supervision, Visualization, Writing - Original Draft
\textbf{Michol A Cooper:} Conceptualization, Data Curation, Project administration, Supervision, Writing - Original Draft  
\textbf{Wei Shao:} Conceptualization, Data Curation, Funding acquisition, Investigation, Methodology, Project administration, Resources, Supervision, Writing - Original Draft

\section*{Conflict of interest statement}
The authors have no conflict of interest to declare.

\section*{Declaration of generative AI and AI-assisted technologies in the writing process}
During the preparation of this work the authors used the ChatGPT-3.5 and ChatGPT-4.0 tools in order to improve the readability and language of our paper. After using this tool/service, the authors reviewed and edited the content as needed and take full responsibility for the content of the publication.

\section*{Acknowledgments}
This work was supported by the Department of Medicine and the Intelligent Clinical Care Center at the University of Florida College of Medicine.
We would like to express our gratitude to Jessica Kirwan for editing the language of this paper.

\bibliographystyle{model2-names.bst}\biboptions{authoryear}
\bibliography{refs}

\end{document}